\documentclass{appolb}
\usepackage{graphicx}
\usepackage{xspace}
\newcommand{\NNLOJET}{NNLO\protect\scalebox{0.8}{JET}\xspace}

\def\pt{$p_{T}$~}

\begin{document}
\title{Single jet inclusive production for the individual jet $p_{T}$ scale choice at the LHC%
\thanks{Presented at the XXIII Cracow Epiphany Conference}%
}
\author{James Currie, E.W.N. Glover
\address{Institute for Particle Physics Phenomenology, Department of Physics, University of Durham,
Durham, DH1 3LE, UK}
\\ \ \\
Thomas Gehrmann
\address{Department of Physics, Universit\"at Z\"urich, Winterthurerstrasse 190,
CH-8057 Z\"urich, Switzerland}
\\ \ \\
Aude Gehrmann-de Ridder, Alexander Huss
\address{Institute for Theoretical Physics, ETH, CH-8093 Z\"urich, Switzerland}
\\ \ \\
Jo\~ao Pires
\address{Max-Planck-Institut f\"ur Physik, F\"ohringer Ring 6 D-80805 Munich, Germany}
}
\maketitle
\begin{abstract}
We study the single jet inclusive cross section up to next-to-next-to leading order
in perturbative QCD, implemented in the parton-level event generator \NNLOJET .
Our results are fully differential in the jet transverse momentum and rapidity
and we apply fiducial cuts for comparison with the available ATLAS 7~TeV 4.5 fb$^{-1}$ data
for jet radius $R=0.4$. For the theoretical calculation we employ the antenna subtraction method
to reliably cancel all infrared divergences present at intermediate stages of the
calculation. We present all results using the individual jet transverse momentum $\mu_{R}=\mu_{F}\sim p_{T}$
as the renormalization and factorization scale for each jet's contribution to the single jet inclusive cross section.
Finally, we consider the differences between our predictions using this scale choice to
those for the leading jet transverse momentum scale choice, $\mu_{R}=\mu_{F}\sim p_{T_{1}}$, used in~\cite{Currie:2016bfm}, 
with reference to the ATLAS data.

\end{abstract}
  
\section{Introduction}

\begin{figure}[htb]
\centerline{%
\includegraphics[width=10cm]{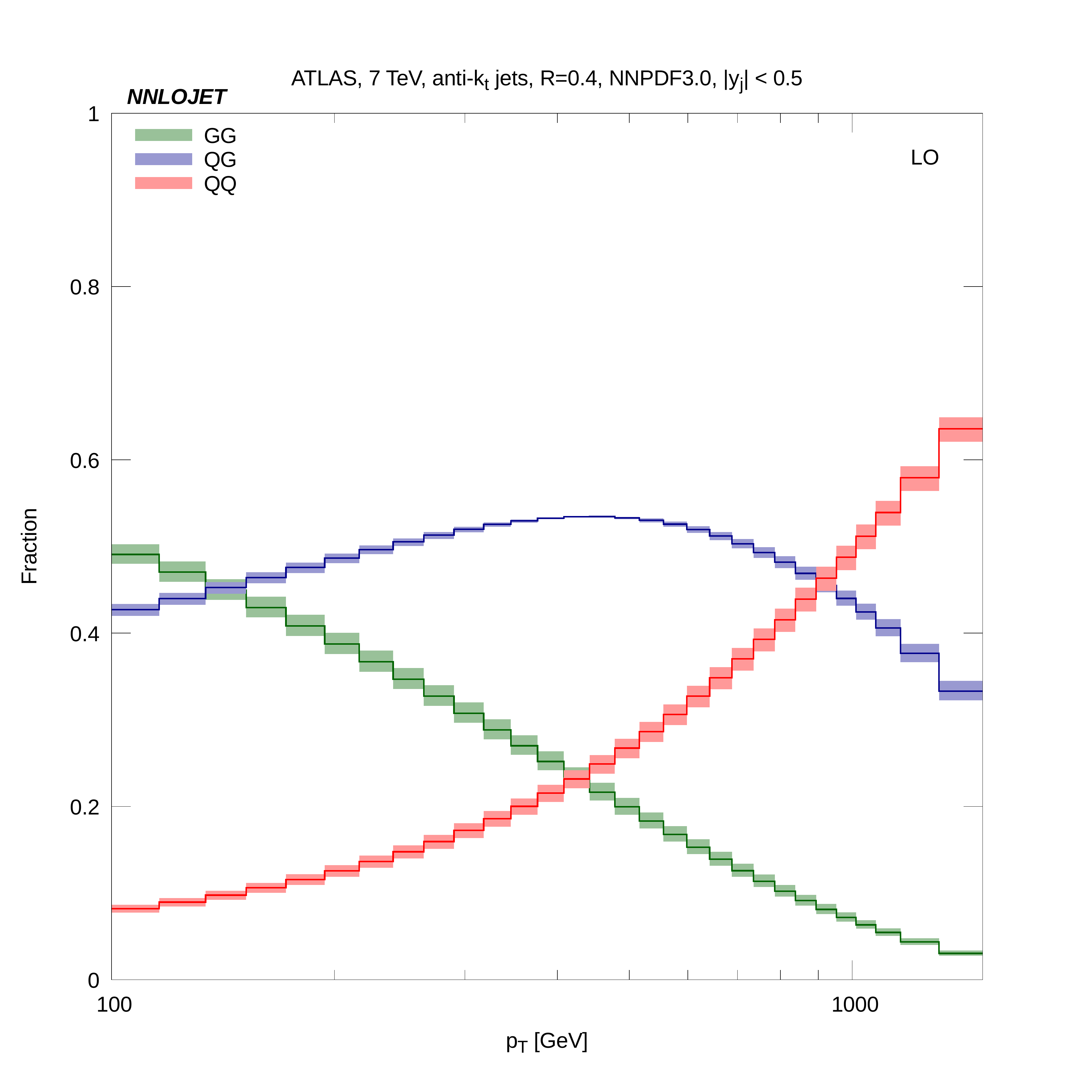}}
\caption{The fraction of jets associated with a given initial-state for inclusive jet production calculated at
LO for the LHC at 7~TeV. The different bands
denote different initial-states: $gg$ (green), $qg$ (blue), $qq$ (red). The relative size
of the contributions goes from $gg>qg>qq$ at low $p_{T}$ to an inverted hierarchy $qq>qg>gg$ at high $p_{T}$.}
\label{Fig:frac}
\end{figure}

The single jet inclusive cross section is one of the most basic observables at any hadron collider.
At its heart is the $2\to2$ QCD subprocess which already at leading order (LO) carries
a dependence on the strong coupling, $\alpha_{s}$, and significant sensitivity to the parameterization and
value of the gluon's Parton Distribution Function (PDF), 
as can be seen in Fig.~\ref{Fig:frac}. It is clear that the $gg$ and $qg$ initial states dominate the 
production of jets over much of the experimentally accessible range in transverse momentum.

This observable has been measured by the ATLAS~\cite{atlasjet} and CMS~\cite{cmsjet} experiments at the LHC
and has been used for determinations of $\alpha_{s}$~\cite{Malaescu:2012ts,Khachatryan:2014waa} and in global 
PDF fits~\cite{NNPDF,MMHT,CT14}. These studies have so far been carried out using theoretical predictions
at next-to-leading order (NLO) accuracy in the strong coupling~\cite{nlojet1,powheg2j,meks} which is typically
accurate at the 10\% level (although this can be higher or lower in specific regions of phase space). Where jet
data has been included in NNLO PDFs this has been done using approximate NNLO predictions based upon
threshold resummation tecnniques~\cite{threshold1,threshold2}.

The single jet inclusive cross section approximately scales with the jet transverse energy, $E_{T}$, as,
\begin{eqnarray}
\frac{{\rm d}^2\sigma}{{\rm d}p_{T}{\rm{d}}y}\sim E_{T}^{-5}.
\end{eqnarray}
The dominant systematic error in the measurement of jets comes from the Jet Energy Scale (JES)
~\cite{Aad:2014bia} which is typically at the level of 1-2\% for central jets at moderate $p_{T}$
 (but significantly larger for very high \pt and forward rapidity bins). This uncertainty in the jet energy translates into a 
systematic uncertainty on the cross section of $\sim$~5-10\%, as is confirmed by the 
detailed quantitative study of such errors in~\cite{atlasjet}. 

To improve the theoretical description of the single jet inclusive cross section in line with experimental advances,
we have recently reported the calculation of the next-to-next-to leading order (NNLO) correction to jet production~\cite{Currie:2016bfm}.
The inclusion of the NNLO contribution should systematically reduce the theoretical uncertainty as estimated by the 
magnitude of the variation of the observable upon variation of the unphysical renormalization, $\mu_{R}$, and factorization,
$\mu_{F}$, scales.

\section{Jets at the LHC}

To make a connection between a parton-level calculation in perturbation theory and the experimentally
observable jet found in the detector, it is necessary to employ a jet algorithm to cluster both
the partons of the theoretical calculation and the energy deposits in the calorimeter into jets. The most
commonly used class of jet algorithms for LHC era jet studies are the sequential recombination algorithms.
These algorithms are characterized by a resolution parameter, $R$ and an integer, $p$~\cite{Salam:2009jx}.
 The algorithm begins with the set of parton momenta coming from a point in phase space and sequentially 
 clusters the momenta into ``proto-jets'' using the \emph{distance} measures,
 \begin{eqnarray}
 d_{ij}&=&{\rm min}(p_{T,i}^{2p},p_{T,j}^{2p})\frac{R_{ij}^{2}}{R^2},\ \ \ R_{ij}^2={(y_{i}-y_{j})^2+(\phi_{i}-\phi_{j})^2},\nonumber\\
 d_{iB}&=&p_{T,i}^{2p}.
 \end{eqnarray}
If the smallest distance calculated is a $d_{ij}$ then the two proto-jet momenta are merged using a ``recombination scheme'' into a single
 proto-jet and the algorithm starts over. 
 The standard recombination scheme is the ``4-vector'' scheme which simply
 adds the 4-momenta of the proto-jets and so generically produces massive proto-jets from massless partons.
 If the smallest distance calculated is a $d_{iB}$ then that proto-jet is removed from the list of proto-jets, stored as a ``jet candidate'' and
 the algorithm starts over.
 Once all proto-jets have been iteratively merged or removed and labelled as jet candidates, the algorithm terminates.
 We then apply the fiducial cuts to the set of
 jet candidates and those which survive the cuts are identified as jets which can be compared to data. Different choices of the parameter
 $p$ define different algorithms with the $k_{T}$~\cite{Ellis:1993tq}, Cambridge/Aachen~\cite{Wobisch:1998wt}
  and anti-$k_{T}$~\cite{antiKT} algorithms defined for  $p=1,0$ and -1 respectively.  The value of the resolution
  parameter $R$ defines how far in $y$-$\phi$ space (which provides cylindrical coordinates for the detector geometry)
  the algorithm reaches out to merge proto-jets. For the purposes of comparing to ATLAS data we
  employ the anti-$k_{T}$ jet algorithm with $R=0.4$ throughout this paper. 

\section{Antenna Subtraction}

It is well known that the various contributions to the physical cross section, as calculated
in perturbation theory, contain infrared (IR) singularities, either as explicit poles in the
dimensional regularization parameter $\epsilon=(4-d)/2$ or as unregulated divergences in
the phase space integrals over parton-momenta.

For inclusive jet production at NNLO we have the double real (RR) contribution given by the tree-level six-parton amplitude
squared~\cite{real}, the real-virtual (RV) contribution given by the interference of the one-loop with the tree-level
five parton amplitudes~\cite{fiveg1l,2q3g1l,4q1g1l} and the double virtual (VV) given by the interference of the two-loop with tree-level
and self-interference of the one-loop four-parton amplitudes~\cite{twol,quarkgluon2l,fourq2l}.

These contributions can be integrated numerically in four dimensions by introducing a set of local subtraction terms so as to
reorganize the NNLO partonic cross section for initial-state partons of species $i,j$, into the form,
\begin{eqnarray}
{\rm d}\hat{\sigma}_{ij}^{NNLO}&=&\int {\rm d}{\Phi_{4}}~\bigg[{\rm d}\hat{\sigma}_{ij}^{RR}-{\rm d}\hat{\sigma}_{ij}^{S}\bigg]\nonumber\\
&+&\int {\rm d}{\Phi_{3}}~\bigg[{\rm d}\hat{\sigma}_{ij}^{RV}-{\rm d}\hat{\sigma}_{ij}^{T}\bigg]\nonumber\\
&+&\int {\rm d}{\Phi_{2}}~\bigg[{\rm d}\hat{\sigma}_{ij}^{VV}-{\rm d}\hat{\sigma}_{ij}^{U}\bigg].
\end{eqnarray}
The subtraction terms are constructed from antenna functions~\cite{antenna1,antenna2} and reduced multiplicity matrix elements. The details
of the construction of the various subtraction terms can be found in~\cite{antenna1,Currie:2017tpe}.

\section{Theoretical scale choice}

  The single jet inclusive cross section is accumulated by binning every jet in an event with at least one
  jet according to its transverse momentum, $p_{T}$, and rapidity, $y$. As such, it is important to remember
  that a single event can contribute several times to the distributions. For the theoretical calculation, each jet is 
  binned with a weight which depends on the value of the appropriate PDF and $\alpha_{s}$, which in turn
  depend on the chosen values of the theoretical scales $\mu_{F}$ and $\mu_{R}$ respectively. The fact that
  contributions to inclusive distributions come from individual jets, rather than events, introduces an
  ambiguity to the choice of theoretical scale; should we set the 
  theoretical scales to reflect the hardness of the individual jets or the event from which they originated?
  
  In a previous study~\cite{Currie:2016bfm} we set the theoretical scales equal to the transverse momentum of the 
  hardest jet in the event, denoted $p_{T_{1}}$. This is an event-wide scale choice and is applied to the weights
  carried by \emph{all} jets in an event, such that in a four jet event, the value of $\alpha_{s}$ and the PDF
  weight is the same for the contribution of the fourth jet as it is for the leading jet.
  
  An alternative is to use the individual jet \pt as the theoretical scale for each jet entering the distribution. For
  the leading jet in the event this scale is identical to $p_{T_{1}}$ and so 1-jet events, where only a single
  jet survives the fiducial cuts, is insensitive to the scale choice between $p_{T}$ and $p_{T_{1}}$. Similarly,
  2-jet events where the jets are balanced in $p_{T}$ cannot generate any difference 
  as $p_{T}=p_{T_{1}}=p_{T_{2}}$. Away from these jet configurations, the 
  subleading jets will have smaller \pt than the leading jet in the event and so choosing the individual
  jet \pt as the theoretical scale will mean that the scale used to calculate the
  weight associated with a jet will on average be smaller than the scale $p_{T_{1}}$.
  A smaller value of $\mu_{R}$ will induce a larger value of $\alpha_{s}$ for the subleading jets 
  and a smaller value of $\mu_{F}$ will alter the relative values of the PDFs. These differences will
  affect the calculated cross section in regions of phase space where the \pt of the subleading jets differs significantly
  from $p_{T_{1}}$, particularly for jets with low transverse momentum.

\section{Results for $\mu_{R},\mu_{F}\sim p_{T}$}

The results presented here are for the experimental setup (\pt and rapidity bin widths) used by the ATLAS collaboration
for the $\sqrt{s}=7$~TeV 4.5 fb$^{-1}$ data set with jets reconstructed using the anti-$k_{T}$ jet algorithm with $R=0.4$. 
The cuts imposed on the jet data include all jets found with $p_{T}\ge 100$~GeV
and $|y|<3$. The theoretical calculation uses the NNPDF3.0 NNLO PDF set with $\alpha_{s}(M_{Z}^2)=0.118$ for LO, NLO
and NNLO contributions. The unphysical theoretical scales are set equal to the individual jet $p_{T}$ such that for a jet
with transverse momentum $p_{T}$, $\mu_{R}=\mu_{F}=p_{T}$ is the central scale choice. To obtain an estimate of the residual theoretical
uncertainty associated with the unphysical scales we vary the central scale choice by factors of two and one half to obtain
an envelope of predictions for the cross section.

\begin{figure}[t]
\centerline{%
\includegraphics[width=10cm]{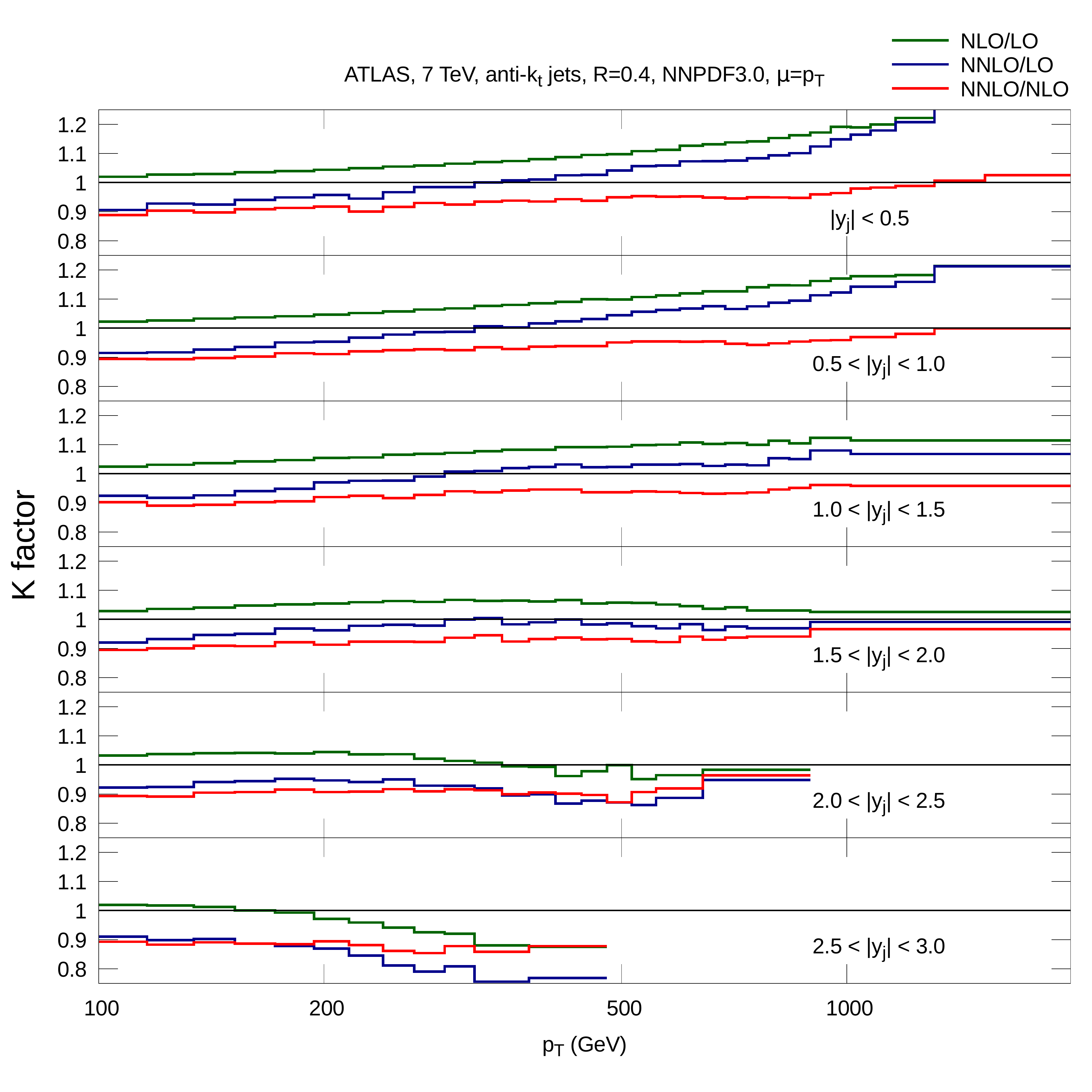}}
\caption{NLO/LO (green), NNLO/NLO (red) and NNLO/LO (blue) $K$-factors for jet production at $\sqrt{s}=7$ TeV. The lines correspond to the double differential
   $K$-factors (ratios of perturbative predictions
  in the perturbative expansion) for $p_T > 100$~GeV and across six rapidity $|y|$ slices. Lines correspond to theoretical predictions evaluated with NNLO PDFs
  from NNPDF3.0 and central scale choice $\mu_{R}=\mu_{F}=p_{T}$.}
\label{Fig:kfac}
\end{figure}

In Fig.~\ref{Fig:kfac} we show the NLO/LO, NNLO/NLO and NNLO/LO $K$-factors across a range of \pt and rapidity bins.
For the central rapidity bin the NLO/LO $K$-factor is small and positive at low \pt and grows to $\sim20\%$ at 1~TeV. In contrast, the NNLO/NLO
$K$-factor provides a negative $\sim10\%$ correction at low \pt and decreases in magnitude at higher $p_{T}$. The overall
behaviour of the higher order corrections is encapsulated in the NNLO/LO $K$-factor which is driven by the NNLO correction at low \pt
and the NLO correction at high $p_{T}$.
As the rapidity of the jets increases we see that the low \pt $K$-factors are similar to those of the central bin. At high \pt the
NLO/LO $K$-factor varies from a large positive correction in the central bin to a moderate negative correction in the most forward rapidity bin.

\begin{figure}[t]
\center{%
(a)\includegraphics[width=5.7cm]{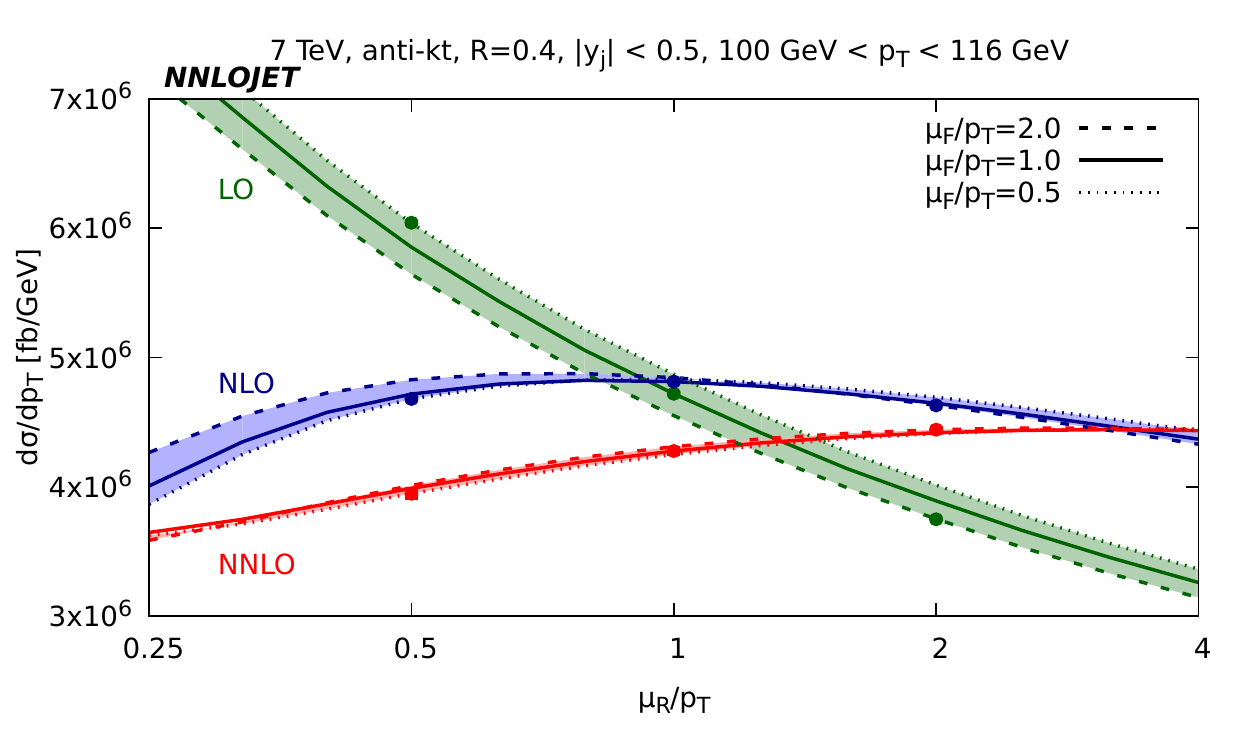}
(b)\includegraphics[width=5.7cm]{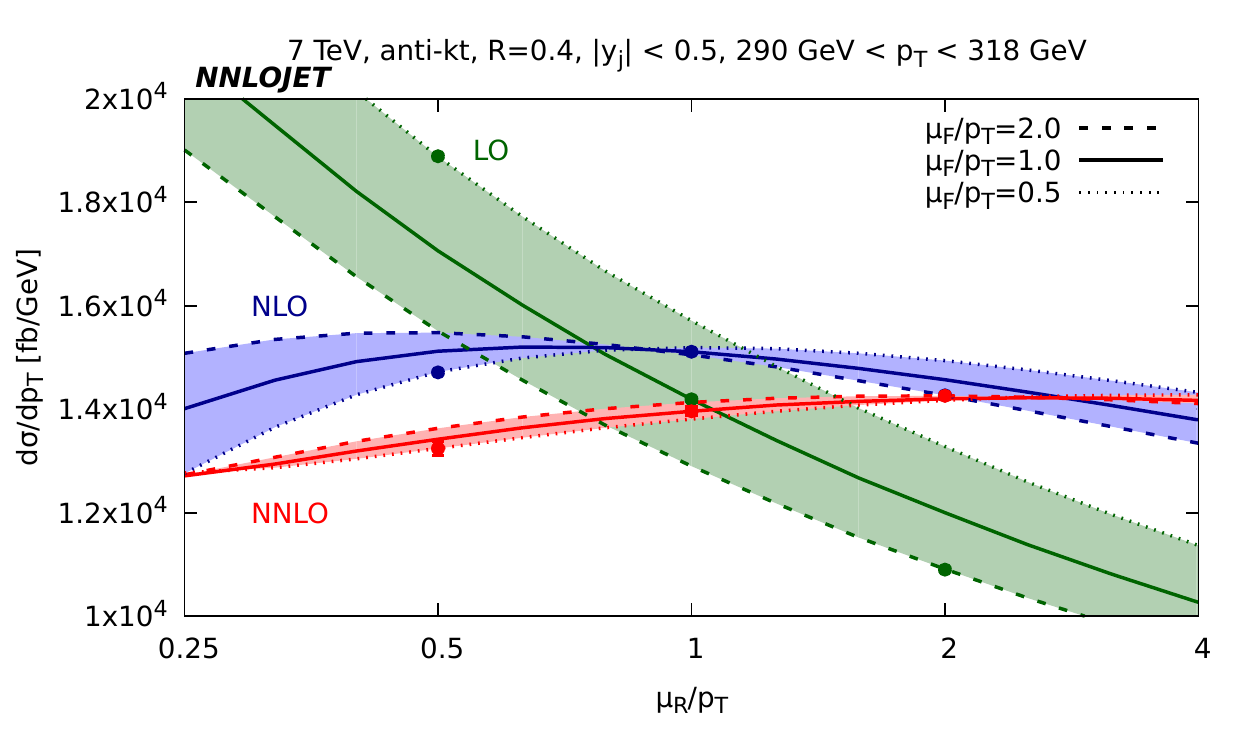}
(c)\includegraphics[width=5.7cm]{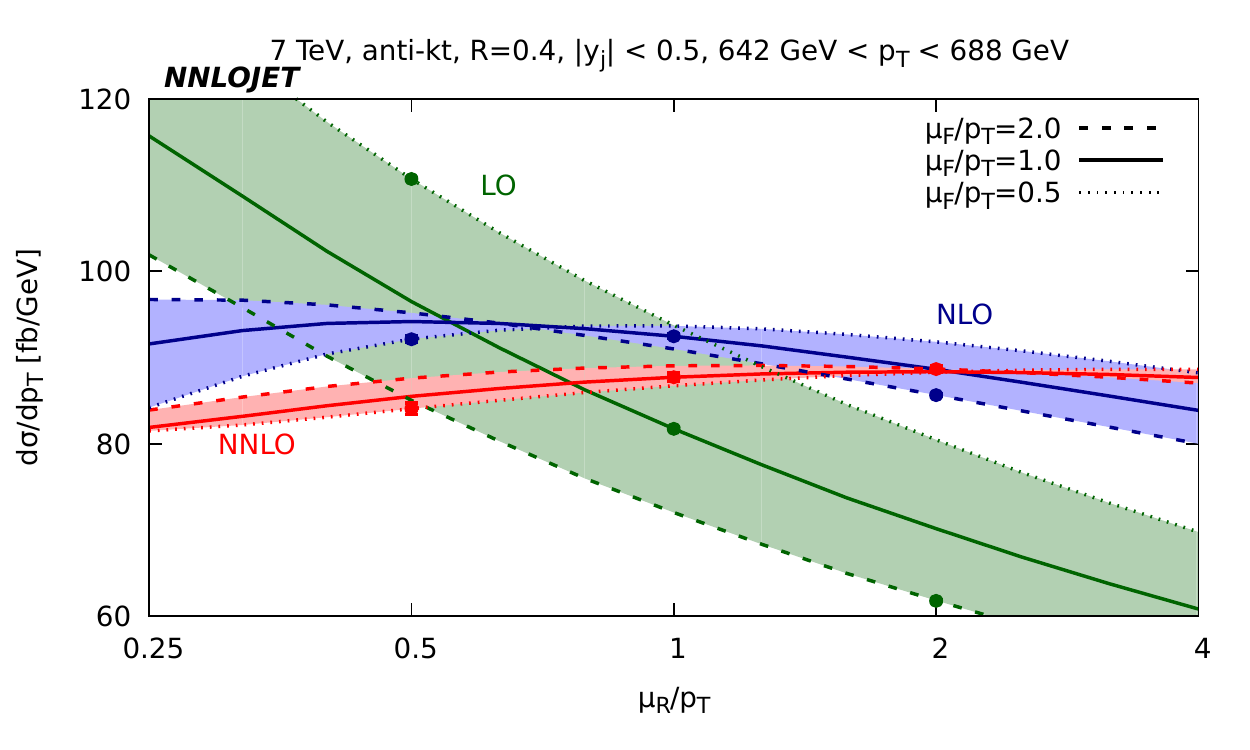}}
\caption{The scale variation of the cross section at LO (green), NLO (blue) and NNLO (red) for central rapidity and three different
\pt bins: (a) 100~GeV$<p_{T}<$116~GeV, (b) 290~GeV$<p_{T}<$318~GeV, (c) 642~GeV$<p_{T}<$688~GeV.}
\label{Fig:scale}
\end{figure}

In addition to the size and shape of the theoretical predictions, the NNLO contribution can affect the residual scale variation. In Fig.~\ref{Fig:scale}
we show how the cross section changes upon variation of the renormalization and factorization scales in a low, medium and high \pt bin for central rapidity.
Across all bins we observe that for fixed $\mu_{F}$ the LO cross section varies monotonically with the variation of the renormalization scale, as is to be expected as
$\mu_{R}$ only affects the value of $\alpha_s$ at LO. 

At NLO we observe a more complicated variation due to the appearance of scale logarithms in the calculation
which can oppose the variation coming from the strong coupling for $\mu_{R},\mu_{F}<\mu_{0}$, where $\mu_{0}$ is the central scale choice, i.e. $p_{T}$.
 The resultant shape has a maximum
in the region of $\mu_{R}/p_{T}\sim$0.5-1 depending on the \pt bin. The peaked shape of NLO curve ensures that the variation of the cross section due to
 $\mu_{R}$ is always negative compared to the central value. The position of the peak being close to the central scale choice means that the scale band
 is smallest about this point.
 Whilst such a band gives a true account for the range of values taken by
the cross section upon variation, it gives a misleading estimate of the degree to which the cross section is changing in response to the scale variation.

At NNLO we observe that the curve has less curvature than the NLO curve and is approximately linear with a decreasing gradient for increasing $p_{T}$. 
The variation of the NNLO cross section due to $\mu_{R}$ is larger than NLO
in the low \pt bin, largely owing to the fact that the peaked shape of the NLO curve probably underestimates the uncertainty; 
but even taking this into account, the magnitude of the variation is similar to that at NLO.
At higher \pt the $\mu_{R}$ scale variation of the NNLO cross section decreases as the curve flattens in Fig.~\ref{Fig:scale}(c). At low \pt the change due to
$\mu_{F}$ variation, displayed as the thickness of the bands in Fig.~\ref{Fig:scale}, is relatively small, even at LO;
whereas at high \pt the $\mu_{F}$ variation becomes large at LO and is significantly reduced by including the 
NLO and especially NNLO corrections.

\begin{figure}[t]
\centerline{%
\includegraphics[width=10cm]{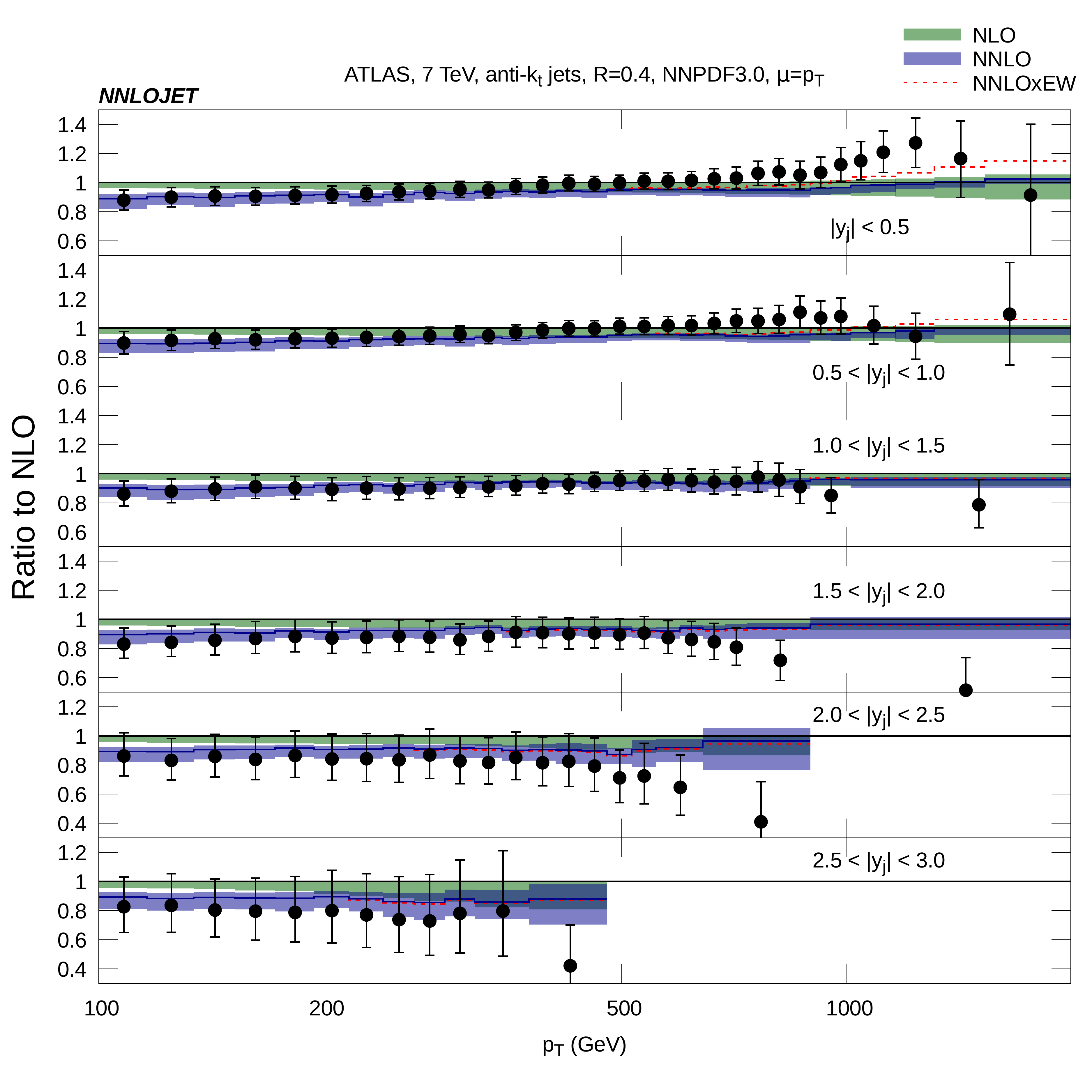}}
\caption{The NLO (green), NNLO (blue) and ATLAS data normalized to the NLO prediction for the individual jet \pt scale choice. The bands 
correspond to the variation of $\mu=\mu_{R}=\mu_{F}$ by factors of 0.5 and 2 about the central scale choice. Electroweak
correction are applied multiplicatively and separately represented as a dashed red line.}
\label{Fig:ratiotonlo}
\end{figure}

The information in Figs.~\ref{Fig:kfac}-\ref{Fig:scale} can be combined and compared to the available ATLAS data, as shown in Fig.~\ref{Fig:ratiotonlo}. 
We observe that at low \pt the NLO prediction shows some tension with the data, which lies approximately 10\% below the NLO prediction. The NNLO
correction acts negatively and brings the theoretical prediction in line with the data. At medium and high \pt the theoretical prediction is largely consistent
with the data and at very high \pt and central rapidity the difference between the NNLO prediction and data can be largely accounted for by including the
NLO electroweak corrections~\cite{eweak}.

\section{Comparison to $\mu_{R},\mu_{F}\sim p_{T_{1}}$}

\begin{figure}[ht]
\centerline{%
 \includegraphics[width=9cm]{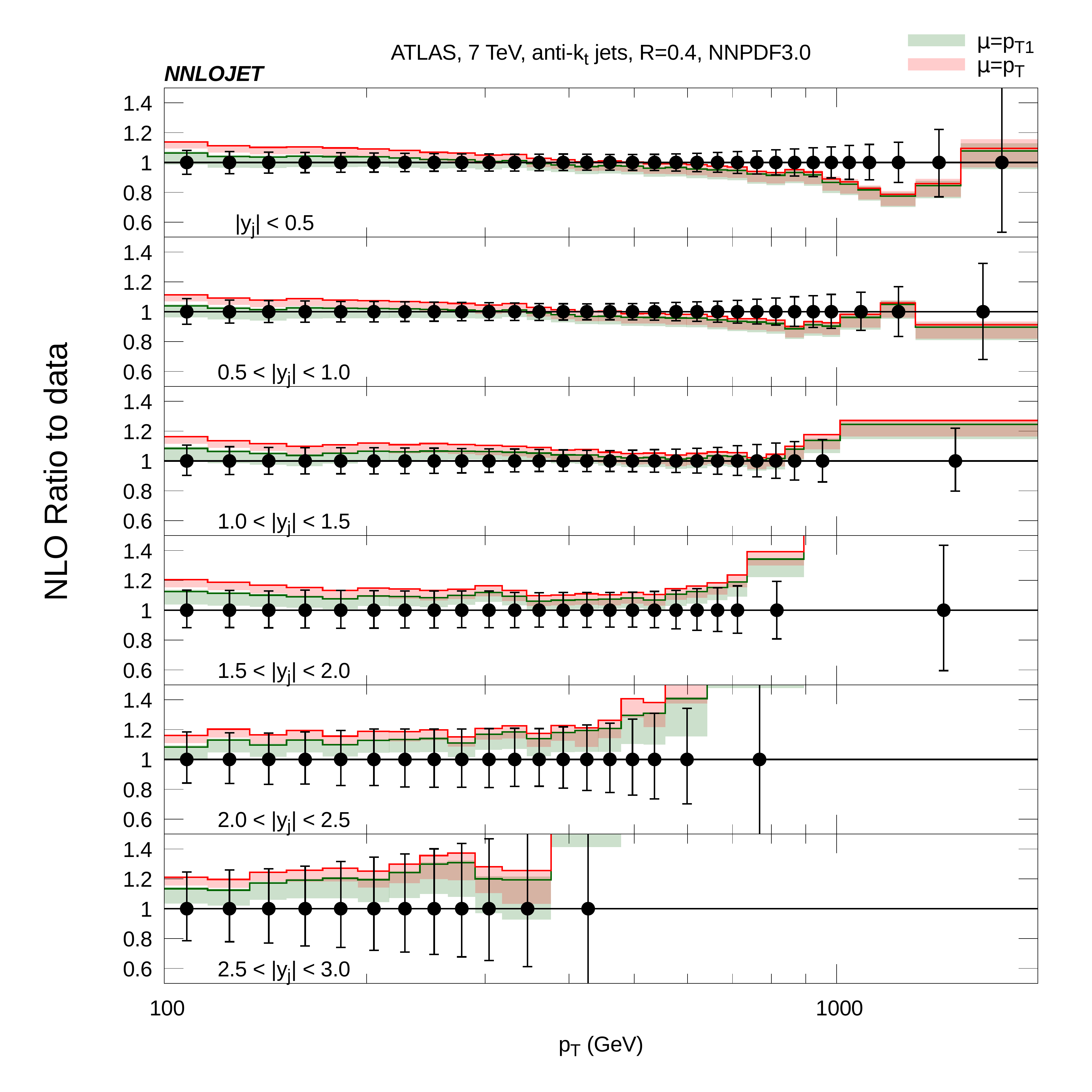}}
\caption{The NLO predictions normalized to data for two different scale choices, individual jet \pt (red) and leading jet \pt (green). The bands 
correspond to the variation of $\mu=\mu_{R}=\mu_{F}$ by factors of 0.5 and 2 about the central scale choice.}
\label{Fig:nlo}
\end{figure}

In addition to comparing to data, we can also compare to the same NNLO calculation, using the leading jet \pt as the theoretical
scale, as reported in~\cite{Currie:2016bfm}.
In Fig.~\ref{Fig:nlo} we show the NLO predictions for difference scale choices, normalized to the data.
 We see that at low \pt there is a significant difference between the 
predictions for the different scale choices with the $p_{T_{1}}$ scale choice sitting close to the data whereas the \pt scale choice lies 
approximately 10\% above the data in the lowest \pt bin for central rapidity. The scale bands at low \pt are of similar size for the two scale choices.
At high \pt we observe that the predictions for the two scale choices converge, particularly for central rapidities. The NLO scale bands are once again
similar in size for the two scale choices and similar in size to the scale bands at low $p_{T}$. 

\begin{figure}[ht]
\centerline{%
\includegraphics[width=9cm]{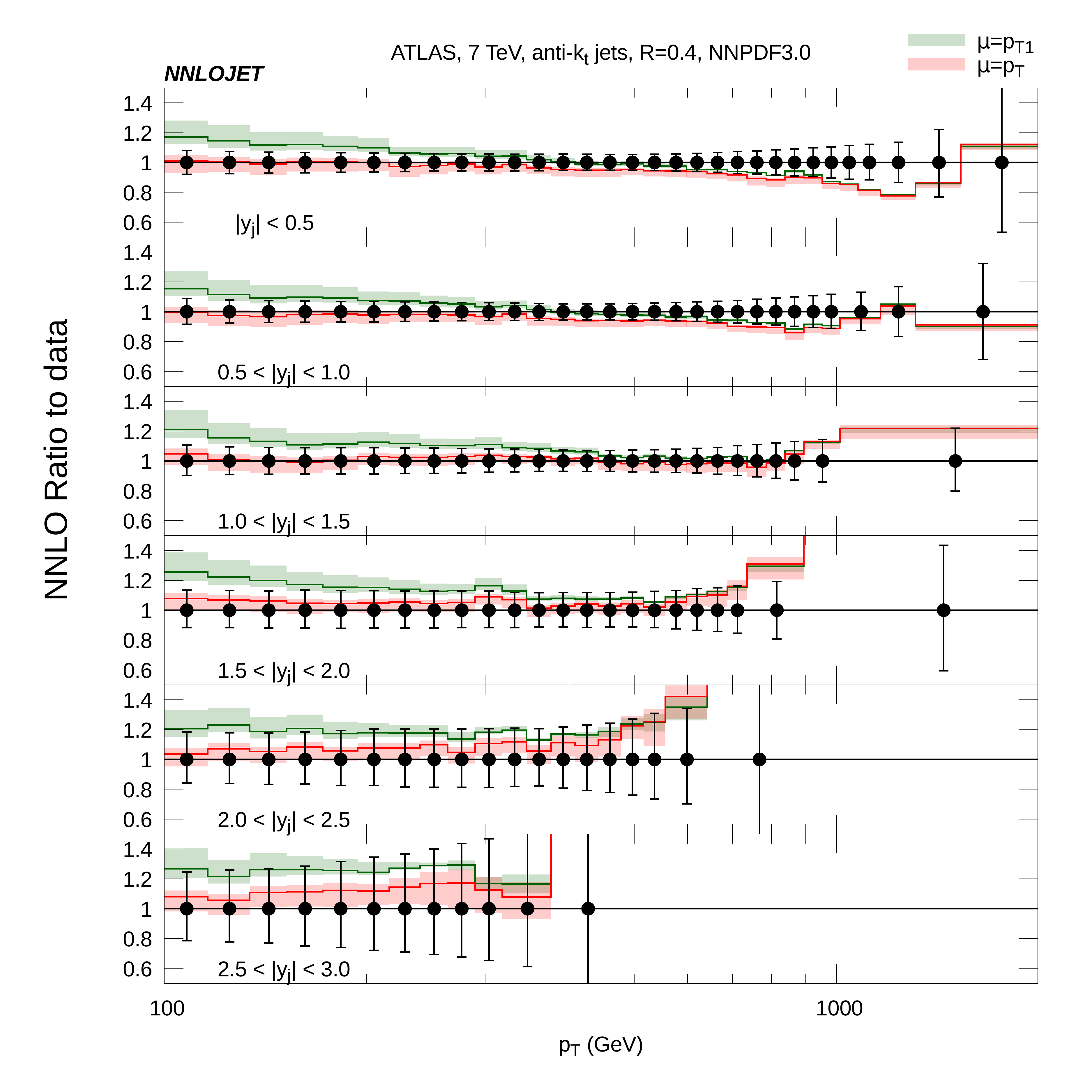}}
\caption{The NNLO predictions normalized to data for two different scale choices, individual jet \pt (red) and leading jet \pt (green). The bands 
correspond to the variation of $\mu=\mu_{R}=\mu_{F}$ by factors of 0.5 and 2 about the central scale choice.}
\label{Fig:nnlo}
\end{figure}

In Fig.~\ref{Fig:nnlo} we show the analogue of Fig.~\ref{Fig:nlo} at NNLO, again for the two scale choices.
At low \pt we find the behaviour somewhat different to NLO: the NNLO correction for the $p_{T_{1}}$ scale moves the prediction away from the data,
with which it was consistent at NLO; whereas using the \pt scale brings the NNLO prediction in line with the data with which there was some tension at NLO.
The NNLO scale band is larger than the NLO scale band for both scale choices in the lowest \pt bin.
At high \pt the predictions for the two scale choices once again converge as is to be expected for the largely back-to-back configurations
found at high $p_{T}$. The NNLO scale band for the \pt scale choice offers only a moderate improvement over the NLO scale band in the \pt range
300-900~GeV (and deterioration below this range) whereas the $p_{T_{1}}$ scale choice shows a more dramatic reduction in scale
uncertainty above 400~GeV and a more dramatic deterioration below 300~GeV. For \pt above 900~GeV both scale choice show a similar
reduction in scale uncertainty when passing from NLO to NNLO predictions.

\section{Discussion}

The individual jet \pt scale choice sets the scale dynamically for each jet in the distribution, rather than at a scale reflecting the hardness
of the event. This scale choice can only produce different results to the previously published results using the leading jet \pt scale choice
when the subleading jets in an event have a \pt differing from the leading jet and this generically occurs
for low \pt jets in the distributions considered here. 

In the low \pt region we find significant differences between the central values for
the predictions using the two scale choices at NLO and NNLO. The uncertainty due to scale variation about those central scales also
increases from NLO to NNLO for both scale choices and despite this increase, the bands do not overlap at low $p_{T}$.
At higher \pt the difference between the two scale choices decreases, as does the uncertainty due to scale variation, although the 
reduction in scale variation is more marked for the $p_{T_{1}}$ scale for moderate \pt and central rapidities.

The comparison to ATLAS data, as shown in Figs.~\ref{Fig:nlo}-\ref{Fig:nnlo}, exemplifies this difference in scale choice. At NLO
we observe that the $p_{T_{1}}$ scale choice sits closer to the data, whereas at NNLO the fortunes are reversed and the \pt
scale choice is more consistent\footnote{The data is being used here merely as a reference point; we are using NNLO PDFs
and so any genuine comparison of the NLO predictions to data is inappropriate. In any case the NLO PDF has been fitted to this data
for the scale choice $p_{T}$.}. 
The inconsistencies between the theoretical calculations clearly poses a problem when it comes to deciding which scale should be used
when comparing to data or fitting PDFs.

The NNLO calculation with \pt scale choice appears to provide a good description of the data, better than with the $p_{T_{1}}$
scale choice. However, it achieves this by generating a relatively large NNLO/NLO $K$-factor alongside a slightly deteriorating 
scale dependence. As an unphysical scale in the theoretical calculation, there is no a priori preferred parameterization except for scales
which minimize the disruptive influence of large logarithms on the perturbative expansion. It is often
sensible to choose a scale which reflects the underlying Born-level kinematics, which for jet production is the LO $2\to2$
scattering where the two scales considered here coincide. The significant effect of this scale ambiguity on the NNLO predictions, and the lack
of a theoretically well motivated preference, motivates further study of this issue and consideration of non-standard scale choices
to ensure the greatest possible phenomenological impact from jet data.

\section*{Acknowledgements}

 The authors thank Xuan Chen, Juan Cruz-Martinez,
  Tom Morgan and Jan Niehues for useful discussions and their many contributions to the \NNLOJET code.
   We gratefully acknowledge the assistance provided by Jeppe Andersen utilizing the computing resources 
   provided by the WLCG through the GridPP Collaboration. 
   This research was supported in part by the UK Science and Technology Facilities Council, in part by the
   Swiss National Science Foundation (SNF) under contracts 200020-162487 and CRSII2-160814, in part by the Research 
   Executive Agency (REA) of the European Union under the Grant Agreement PITN-GA-2012-316704 (``HiggsTools'')
    and the ERC Advanced Grant MC@NNLO (340983).

\end{document}